\documentclass[aps,pra,rexx,twocolumn,superscriptaddress,longbibliography]{revtex4-2}

\usepackage{amsmath}
\usepackage{amssymb}
\usepackage{graphicx}
\usepackage{bm}
\usepackage{xcolor}
\usepackage{mathtools}

\begin{document}

\title{Thermal lifetime of the centralized repetition code: From quantum annealing to social dynamics}

\author{Ari Mizel}
\affiliation{Laboratory for Physical Sciences, College Park, Maryland 20740, USA}

\author{Van Molino}
\affiliation{Center for Computing Sciences, Bowie, Maryland 20715, USA}

\date{\today}

\begin{abstract}
There is an extensive body of research probing the potential of adiabatic quantum computation and quantum annealing to solve hard computational problems.  This research endeavor is complicated by noise afflicting hardware during the computational process. An error-correcting approach called quantum annealing correction (QAC) was suggested to mitigate this noise.  The QAC approach employs a centralized version of the repetition code in which the qubits are configured in a star-graph pattern with one special qubit playing the role of the hub.  In this paper, we explore the thermal physics of the centralized repetition code, using a Lindblad equation framework to analyze its lifetime when coupled to a thermal bath.  We show that its centralized configuration leads to thermal stability, enabling the robust storage of a logical bit of information, despite the fact that there is only 1 ferromagnetic Ising interaction per qubit like a 1-dimensional Ising chain.
\end{abstract}

\maketitle

\section{Introduction}
Adiabatic quantum computation and quantum annealing \cite{Nishimori1998, Farhi2001, Johnson2011,Lidar2018} have been studied intensively in the hope that they will  provide efficient solutions to challenging computational tasks, such as complex optimization problems.  However, physical implementations are susceptible to noise induced by the surrounding environment, such as thermal excitations~\cite{Mizel2002,Childs2001,Albash2015}.  Indeed, while a noise threshold theorem has been proven for gate-model quantum computation \cite{Shor1995, Aharonov2008, Knill1998}, this theorem relies on fast quantum gates and/or measurements and does not apply to the adiabatic quantum computation and quantum annealing context.  In lieu of such a theorem, various mitigations have been suggested to improve the probability of finding the correct answer \cite{Jordon2006,Lidar2008,YoungPRX2013}.

One such mitigation that has attracted attention is a hardware-tailored framework known as quantum annealing correction (QAC) \cite{Pudenz2014}.  This approach involves replacing individual physical qubits within the hardware's unit cells with logical qubits formed out of a repetition code.  In their initial experimental demonstration, Pudenz et al. sought the ground states of antiferromagnetic Ising chains of increasing length~\cite{Pudenz2014}.  Using QAC, they were able to increase appreciably the chances of correctly identifying these ground states on D-Wave hardware.   Following this development, subsequent research further explored benefits of the QAC strategy~\cite{Young2013,Pudenz2015,Vinci2015,Matsuura2016,Pearson2019}.

\begin{figure}
\includegraphics[width=0.5\linewidth]{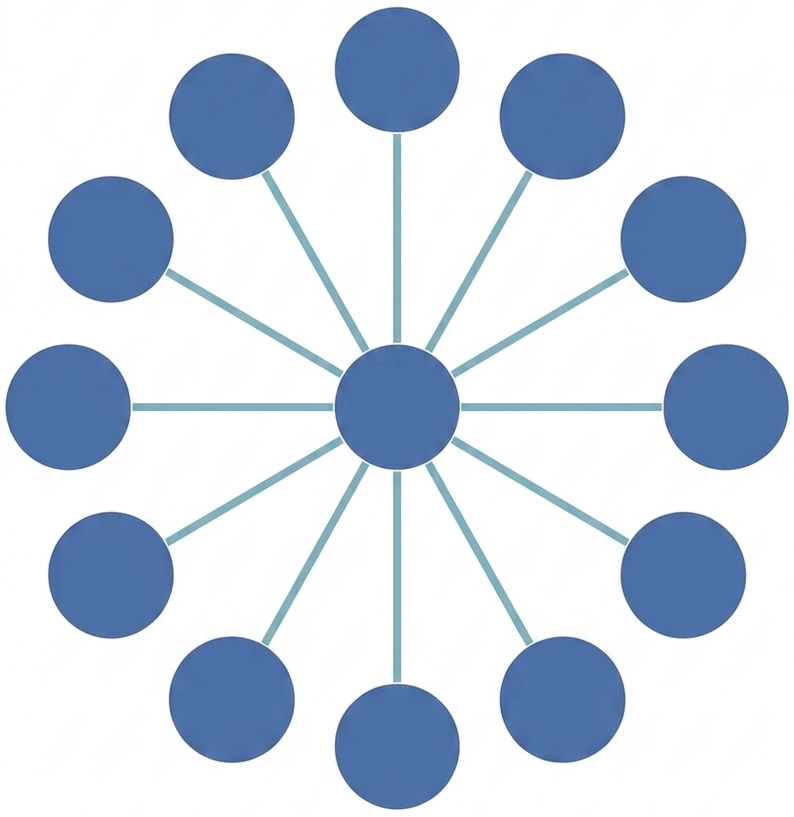}
\caption{Depiction of the centralized repetition code in which each circle represents a qubit.  The spokes represent interactions between the central qubit (hub) and each of the peripheral qubits.  The case of $n=13$ qubits is shown.}
\label{fig:code}
\end{figure}

The logical qubit deployed in QAC adopts a star-graph configuration that we refer to as the \textit{centralized repetition code}.  In a centralized repetition code containing $n$ qubits, a single central physical qubit (the ``hub'') is linked via ferromagnetic Ising interactions (the ``spokes'') to the remaining $n-1$ peripheral qubits.   A schematic is shown in Fig. \ref{fig:code}.
On the one hand, every peripheral qubit in this configuration participates in only a single Ising bond.  In this respect, it is similar to a 1-dimensional ferromagnetic Ising chain.  On the other hand, the high connectivity of the central hub makes it reminiscent of a complete graph in which every qubit connects to every other qubit.  Thus, it is natural to ask whether the system behaves thermodynamically like the former, which cannot store information robustly at any non-zero temperature, or the latter.

In this paper, we investigate the thermal lifetime of the centralized repetition code as a function of the total qubit number $n$. Studying the thermal evolution of the code using a Lindblad equation \cite{Breuer2007, Weiss2012}, we establish that its storage lifetime scales exponentially with system size. Our results indicate that the centralized repetition code is thermally more similar to the repetition code on a complete graph than a 1-dimensional chain.  This is consistent with an analysis of the mixing of discrete time Glauber dynamics on spin systems containing high-degree vertices \cite{Blanca2023}.  

The paper is organized as follows.  In sections II and III, we present the stabilizer Hamiltonian of the code and derive a master equation describing its evolution in contact with a thermal bath.  In section IV, we study the solution of the master equation using spectral analysis.  Section V uses the spectral analysis to characterize the thermal lifetime of the code.  We conclude in Section VI.

\section{Stabilizer Hamiltonian}
We consider an $[n,1,n]$ repetition code consisting of $n$ physical qubits, where $n$ is a positive odd integer. The logical $0$ state is encoded as $|0, 0, \dots, 0\rangle$ and the logical $1$ state as $|1, 1, \dots,1\rangle$. We assign the central physical qubit (the hub) the index 1 and label the peripheral qubits from 2 to $n$. The code's stabilizers  \cite{Gottesman1997} are given by $Z_1 Z_i$ for $i=2,\dots, n$, meaning the first qubit serves as a central standard to which all other qubits are compared. 

The choice of stabilizers yields the stabilizer Hamiltonian
\begin{equation}
H^{\text{stab}} = -J \sum_{i=2}^n Z_1 Z_i,
\end{equation}
where $J>0$ parameterizes the coupling strength and $Z_i$ denotes the standard Pauli $\sigma_Z$ operator acting on the $i$-th qubit. This Hamiltonian has the code words $|0, 0, \dots, 0\rangle$ and $|1, 1, \dots,1\rangle$ as ground states and imposes an energy penalty when qubits flip. The corresponding energy eigenstates are $|b_1,b_2, \dots, b_n\rangle$, where $b_i \in \{0,1\}$, yielding the eigenvalues:
\begin{equation}
E_{b_1, \dots, b_n} = -J (-1)^{b_1} \sum_{i=2}^n (-1)^{b_i}.
\label{eq:E_stabilizer}
\end{equation}

\section{Master Equation}
To quantify its resilience against environmental noise, we couple the code to a thermal bath. The total Hamiltonian of the composite system is modeled as
\begin{equation}
H = H^{\text{stab}} + H^{\text{bath}} + H^{\text{int}}.
\end{equation}
Here, the interaction term is composed of local bit-flip errors acting on each individual qubit
\begin{equation}
H^{\text{int}} = \sum_{i=1}^n X_i \otimes f_i,
\label{eq:Hint}
\end{equation}
where $f_i$ is some operator acting on the Hilbert space of the thermal bath and $H^{\text{bath}}$ governs the evolution of the $f_i$.

Under standard Born-Markov assumptions regarding the bath, the reduced density matrix $\rho$ of the code evolves according to the Lindblad equation \cite{Breuer2007, Weiss2012}:
\begin{equation}
\begin{split}
\frac{d\rho}{dt} = &-i[H^{\text{stab}},\rho] \\
&+\sum_{i,\omega} g_i(\omega) \left(S_i(\omega)\rho S_i^\dagger(\omega) - \frac{1}{2} \{S_i^\dagger(\omega)S_i(\omega),\rho\}\right).
\end{split}
\label{eq:Lindblad}
\end{equation}
The jump operator $S_i(\omega)$ effects a transition in which the system changes its energy by $-\omega$ in accordance with the definition
 $e^{iH^{\text{stab}}t} X_i e^{-iH^{\text{stab}}t} = \sum_\omega S_i(\omega) e^{-i\omega t}$. The transition rates satisfy the detailed balance condition $g_i(-\omega) = e^{-\beta \omega}g_i(\omega)$ where $\beta = 1/k_B T$ is the inverse temperature of the thermal bath.

Because the central qubit $i=1$ plays a special role in the stabilizer Hamiltonian, the form of $S_i(\omega)$ is different for $i>1$ than it is for $i=1$. For peripheral qubits ($i > 1$), flipping a qubit from being opposite to $b_1$ to matching $b_1$ drops the system's energy by $2J$, whereas flipping from matching $b_1$ to being opposite increases it by $2J$. Thus, the non-vanishing jump operators for $i > 1$ are:
\begin{align}
S_i(2J) = \sum_{\mathclap {b_1,\dots,b_{i-1},b_{i+1},\dots,b_n}} & |b_1, \dots, b_i=b_1, \dots, b_n\rangle \nonumber \\
& \hspace{0.2in} \langle b_1, \dots, b_i=1-b_1, \dots, b_n| \\
S_i(-2J) = \sum_{\mathclap {b_1,\dots,b_{i-1},b_{i+1},\dots,b_n}}& |b_1, \dots, b_i=1-b_1, \dots, b_n\rangle \nonumber \\
&  \hspace{0.2in} \langle b_1, \dots, b_i=b_1, \dots, b_n|.
\end{align}
In contrast, flipping the central qubit ($i=1$) alters the system energy by $\pm 2J\ell$, where $\ell = \sum_{k=2}^n (-1)^{b_k}$ is an integer reflecting the net balance of the peripheral spins. The respective jump operators evaluate to:
\begin{align*}
S_1(2J\ell) = &\sum_{b_1,\dots,b_n} \Theta((-1)^{b_1}\ell)\delta_{\ell, \sum_{k=2}^n(-1)^{b_k}} \\
&\times |b_1,\dots,b_n\rangle\langle 1-b_1,\dots,b_n| \\
S_1(-2J\ell) = &\sum_{b_1,\dots,b_n} \Theta(-(-1)^{b_1}\ell)\delta_{\ell, \sum_{k=2}^n(-1)^{b_k}} \\
&\times |b_1,\dots,b_n\rangle\langle 1-b_1,\dots,b_n|
\end{align*}
where $\Theta(k)$ is the standard Heaviside step function
\begin{align*}
\Theta(k) & = \left\{ \begin{array}{ll} 1 & k \ge 0 \\ 0 & k < 0 \end{array} \right. .
\end{align*}

Assume that the initial density matrix of the code is an incoherent mixture of energy eigenstates, $\rho(0) = \sum_{b_1,\dots,b_n} p(b_1,\dots,b_n) |b_1,\dots,b_n\rangle \langle b_1,\dots,b_n|$.  Then, the off-diagonal components of $\rho(t)$ remain zero under the evolution (\ref{eq:Lindblad}) because $[H^{\text{stab}}, \rho] = 0$. Defining the column vector of occupation probabilities 
\begin{align}
\mathbf{p} & = [p(0,\dots,0), \dots, p(1,\dots,1)]^T,
\end{align}
we find that Eq.~\eqref{eq:Lindblad} reduces exactly to a classical master equation:
\begin{equation}
\label{eq:master}
\frac{d\mathbf{p}}{dt} = M \mathbf{p},
\end{equation}
where $M = \sum_{i=1}^n M_i$ represents the sum of transition rate matrices for each individual bit flip. For the peripheral qubits $i > 1$, $M_i$ is a block-diagonal matrix whose $2 \times 2$ blocks in the basis $|b_i\rangle = \{|0\rangle, |1\rangle\}$ take the form:
\begin{equation}
\left[\begin{array}{rr}
-g_i(-2J(-1)^{b_1}) &  g_i(2J(-1)^{b_1}) \\ 
 g_i(-2J(-1)^{b_1}) & -g_i(2J(-1)^{b_1})
\end{array} \right].
\label{eq:Mi}
\end{equation}
Relaxation rates $g_i(2J)$ are related to excitation rates via $g_i(-2J) = g_i(2J)e^{-2\beta J}$. For the central qubit $i=1$, we find the following $2 \times 2$ blocks in the basis $|b_1\rangle = \{|0\rangle, |1\rangle\}$:
\begin{equation}
\left[\begin{array}{rr}
-g_1\left(-2J\sum_{k=2}^n(-1)^{b_k}\right) &  g_1\left(2J\sum_{k=2}^n(-1)^{b_k}\right) \\ 
 g_1\left(-2J\sum_{k=2}^n(-1)^{b_k}\right) & -g_1\left(2J\sum_{k=2}^n(-1)^{b_k}\right)
\end{array} \right].
\end{equation}

\section{Solving the Master Equation with Spectral Analysis}

To analyze the master equation Eq.~\eqref{eq:master}, we employ spectral analysis.  The eigenvalues and eigenvectors of $M$ are defined by
\begin{equation}
M {\bf p}_k = \lambda_k {\bf p}_k.
\end{equation}
The standard equilibrium Boltzmann distribution $\mathbf{p}_\beta$, with components given by
\begin{equation}
p_\beta(b_1,\dots,b_n) = \frac{e^{\beta J (-1)^{b_1} \sum_{i=2}^n (-1)^{b_i}}}{\sum_{b_1,\dots,b_n}e^{\beta J (-1)^{b_1} \sum_{i=2}^n (-1)^{b_i}}},
\end{equation}
is an exact eigenvector $\mathbf{p}_0 = \mathbf{p}_\beta$ with eigenvalue $\lambda_0 = 0$.

To rigorously verify the spectral characteristics of the system, we can define a diagonal matrix $P_\beta$ containing the elements of $\mathbf{p}_\beta$ along its diagonal. Transforming the rate matrices into $\tilde{M} = P_\beta^{-1/2} M P_\beta^{1/2}$ yields a Hermitian matrix.  Indeed, the individual matrix $\tilde{M}_i = P_\beta^{-1/2} M_i P_\beta^{1/2}$ is Hermitian because detailed balance enforces 
\begin{align*}
g_i(E_a-E_b) &e^{\beta E_b/2} e^{-\beta E_a/2} \\ 
&= g_i(E_b-E_a) e^{\beta E_a/2} e^{-\beta E_b/2}.
\end{align*}
Consequently, the total matrix $\tilde{M} = \sum_i \tilde{M}_i$ is Hermitian.

Because $M$ has the same spectrum as the Hermitian matrix $\tilde{M}$, all eigenvalues $\lambda_k$ are guaranteed to be real numbers. Furthermore, a direct application of the Gershgorin circle theorem \cite{Gershgorin1931} to the columns (which sum to zero to ensure trace preservation of ${\bf p}$) proves that the eigenvalues of $M_i$ are negative semi-definite, satisfying $\lambda_k \le 0$.

The matrix $M$ possesses a second eigenvalue $\lambda_1$ that lies close to zero. We show this with a variational argument.  To construct an ansatz for the second eigenvector, we define the logical-Z value of the basis vector $|b_1,\dots,b_n\rangle$ to be $(-1)^{\text{majority}(b_1,\dots,b_n)}$.  We arrange these logical-Z values into the components of a vector ${\bf z}$
\begin{equation}
z_{b_1,\dots,b_n} = (-1)^{\text{majority}(b_1,\dots,b_n)}.
\end{equation}
Then, we define a logical-Z matrix ${\cal Z}$ that has ${\bf z}$ on its diagonal and vanishes elsewhere.
An excellent ansatz ${\bf p}^a$ for the second eigenvector is obtained by weighting the Boltzmann distribution by this logical-Z, ${\bf p}^a = {\cal Z} {\bf p}_\beta$.  Here, the superscript $a$ stands for ansatz.  This guess was inspired by the analysis in \cite{Alicki2009,Alicki2010}.  The components of ${\bf p}^a$ are given by
\begin{equation}
p^a_{b_1,\dots,b_n} = \frac{(-1)^{\text{majority}(b_1,\dots,b_n)} e^{-\beta E_{b_1,\dots,b_n}}}{\sum_{b_1,\dots,b_n}e^{-\beta E_{b_1,\dots,b_n}}}.
\end{equation}
A bound on the second eigenvalue $\lambda_1$ is then obtained using $P_\beta^{-1/2} {\bf p}^a$ as a variational eigenvector of $\tilde{M}$.  To make this argument, we note that $P_\beta^{-1/2} {\bf p}^a$ is orthogonal to $P_\beta^{-1/2}\mathbf{p}_0 = P_\beta^{-1/2}\mathbf{p}_\beta$.  The bound is
\begin{widetext}
\begin{align}
-\lambda_1 & \le - \frac{({\bf p}^a)^T P_\beta^{-1/2} \tilde{M} P_\beta^{-1/2} {\bf p}^a}{({\bf p}^a)^T P_\beta^{-1/2} P_\beta^{-1/2} {\bf p}^a} = - [1,\dots,1] {\cal Z}  M {\cal Z} {\bf p}_\beta  = 4[1 \, \dots \, 1]  \left(\frac{I-{\cal Z}}{2}\right)(M_1+\sum_{i>1}M_i) \left(\frac{I+{\cal Z}}{2}\right) {\bf p}_\beta  \nonumber \\
& = 4 \frac{\splitfrac{\splitfrac{\sum_{b_2,\dots,b_n \text{ half zero}} g_1(E_{b_1=0,\dots,b_n}-E_{b_1=1,\dots,b_n})\,\, e^{-\beta E_{b_1=0,\dots,b_n}}}{+ \sum_{b_1,b_3\dots,b_n \text{ half zero}}g_2(E_{b_1,b_2=0,\dots,b_n}-E_{b_1,b_2=1,\dots,b_n})\,\, e^{-\beta E_{b_1,b_2=0,\dots,b_n}}}}{\dots +\sum_{b_1,\dots,b_{n-1} \text{ half zero}}g_n(E_{b_1,\dots,b_{n-1},b_n = 0}-E_{b_1,\dots,b_{n-1},b_n = 1})\,\, e^{-\beta E_{b_1,\dots,b_{n-1},b_n = 0}}}} { \sum_{b_1,\dots,b_n} e^{-\beta E_{b_1,\dots,b_n}}}   \nonumber \\
 & = 4 \frac{g_1(0)  \binom{n-1}{\frac{n-1}{2}} + \sum_{i>0} (g_i(2J)e^{-2\beta J}  +g_i(-2J))   \binom{n-2}{\frac{n-1}{2}}}{2( e^{\beta J}+e^{-\beta J})^{n-1}}  \le 2 g_{max} \frac{2^{n-1}(1+(n-1) e^{-2\beta J})}{ ( e^{\beta J}+e^{-\beta J})^{n-1}} \nonumber \\
 & \le g_{max} \frac{2(1+(n-1) e^{-2\beta J})}{\cosh^{n-1}(\beta J)} = - \lambda_{\text{bound}} \label{eq:bound}
\end{align}
\end{widetext}
where $g_{max} = \text{max}_{i,\omega} g_i(\omega)$ is assumed independent of $n$.  In the final equality on the first line, we use the fact that $[1,\dots,1]M_i = 0$ and the fact that $M_i {\bf p}_\beta = 0$.  The notation $\sum_{b_2,\dots,b_n \text{ half zero}}$ means that in the sum half of the elements of the set $\{b_2,\dots,b_n\}$ have value $0$ and half have value $1$.  There are $\binom{n-1}{\frac{n-1}{2}} < 2^{n-1}$ ways that this can happen.  Note that, for $\beta J > 0$, the bound (\ref{eq:bound}) goes to zero exponentially in $n$.

The two eigenvectors give rise to two distinct solutions to the master equation ${\bf p}_\beta$ and $e^{\lambda_1 t} {\bf p}_1$. The first solution is stable indefinitely.  The second solution is stable for a characteristic time scale $\tau \sim 1/|\lambda_1|$, which is long since $\lambda_1$ is close to 0.  A solution for which the physical qubits are mostly 0 is ${\bf p}_\beta + e^{\lambda_1 t} {\bf p}_1$.  We approximate this as
\begin{equation}
\begin{split}
p_{b_1,\dots,b_n}(t) \approx &\left(1 + e^{\lambda_{\text{bound}} t} (-1)^{\text{majority}(b_1,\dots,b_n)}\right) \\
&\times \frac{e^{-\beta E_{b_1,\dots,b_n}}}{\sum_{b_1,\dots,b_n}e^{-\beta E_{b_1,\dots,b_n}}};
\end{split}
\end{equation}
this stores logical 0 for $\tau \sim 1/|\lambda_{\text{bound}}|$  before it eventually decays into the trivial Boltzmann distribution.
The corresponding solution for which most physical qubits are 1 is ${\bf p}_\beta - e^{\lambda_1 t} {\bf p}_1$.  We approximate this as
\begin{equation}
\begin{split}
p_{b_1,\dots,b_n}(t) \approx &\left(1- e^{\lambda_{\text{bound}} t}(-1)^{\text{majority}(b_1,\dots,b_n)}\right) \\
&\times \frac{e^{-\beta E_{b_1,\dots,b_n}}}{\sum_{b_1,\dots,b_n}e^{-\beta E_{b_1,\dots,b_n}}},
\end{split}
\end{equation}
which stores logical 1.

One would like an estimate of the error of the variational guess ${\bf p}^a = {\cal Z} {\bf p}_\beta$.   This can be obtained by bounding the next eigenvalue $\lambda_2$.  There is a gap below which $\lambda_2$ stays, irrespective of $n$.  To argue this, consider the eigenvalues of $\tilde{M}_2+\dots+\tilde{M}_n$.  Since $0 = [M_i,M_j] = [\tilde{M}_i,\tilde{M}_j]$ for $i,j> 1$, the eigenvalues are easily obtained.  From the form of (\ref{eq:Mi}), we see that $\tilde{M}_2+\dots+\tilde{M}_n$ has two zero eigenvalues, one for each value of $b_1$.  Its third eigenvalue is 
\begin{align}
-G  & = - \text{ min}_{i>1} \,\, g_i(2J)+g_i(-2J) \nonumber \\
& = \text{ min}_{i>1} \,\,g_i(2J) \,\, (-1 - e^{-\beta 2J}).
\end{align}
Now, we add back in $\tilde{M}_1$ and consider the eigenvalues of $\tilde{M}=\tilde{M}_1+\dots+\tilde{M}_n$.  
Since $\tilde{M}_1$ is negative semi-definite, when we add it back in, $\lambda_2$ remains at least as negative as $-G$.  Expanding $P_\beta^{-1/2} {\cal Z} {\bf p}_\beta$ in a basis of eigenstates $P_\beta^{-1/2}  {\bf p}_j$ of $\tilde{M}$, we write 
$P_\beta^{-1/2} {\cal Z} {\bf p}_\beta = \sum_{j = 1}^{2^n-1} a_j P_\beta^{-1/2}  {\bf p}_j$.  This implies
\begin{align}
& - \frac{({\bf p}^a)^T P_\beta^{-1/2} \tilde{M} P_\beta^{-1/2} {\bf p}^a}{({\bf p}^a)^T P_\beta^{-1/2} P_\beta^{-1/2} {\bf p}^a}  = -\sum_{j = 1}^{2^n-1} |a_j|^2 \lambda_j \nonumber \\
 & \ge - \lambda_1 |a_1|^2 - \lambda_2 (1-|a_1|^2) \ge G (1-|a_1|^2) .
\end{align}
Thus, using the bound (\ref{eq:bound}), the error is limited to
\begin{align}
1-|a_1|^2 \le \frac{g_{max}}{G} \frac{2(1+(n-1) e^{-2\beta J})}{\cosh^{n-1}(\beta J)}
\end{align}
which vanishes exponentially for large $n$.

In the region where (\ref{eq:bound}) is large, on the other hand, the error $1-|a_1|^2$ in our variational guess can be large. Indeed, if $\beta J$ is small, the bound (\ref{eq:bound}) increases for small values of $n$, reaching a maximum at $n = 1 - e^{2\beta J} + 1/\ln \cosh \beta J$
before decreasing with $n$.  Exact numerical diagonalization for small $n$ shows that this behavior is an artifact of the variational approach; the true eigenvalue $\lambda_1$ actually monotonically approaches 0 with increasing $n$. 

\section{Thermal Lifetime}

\begin{figure}
\includegraphics[width=\linewidth]{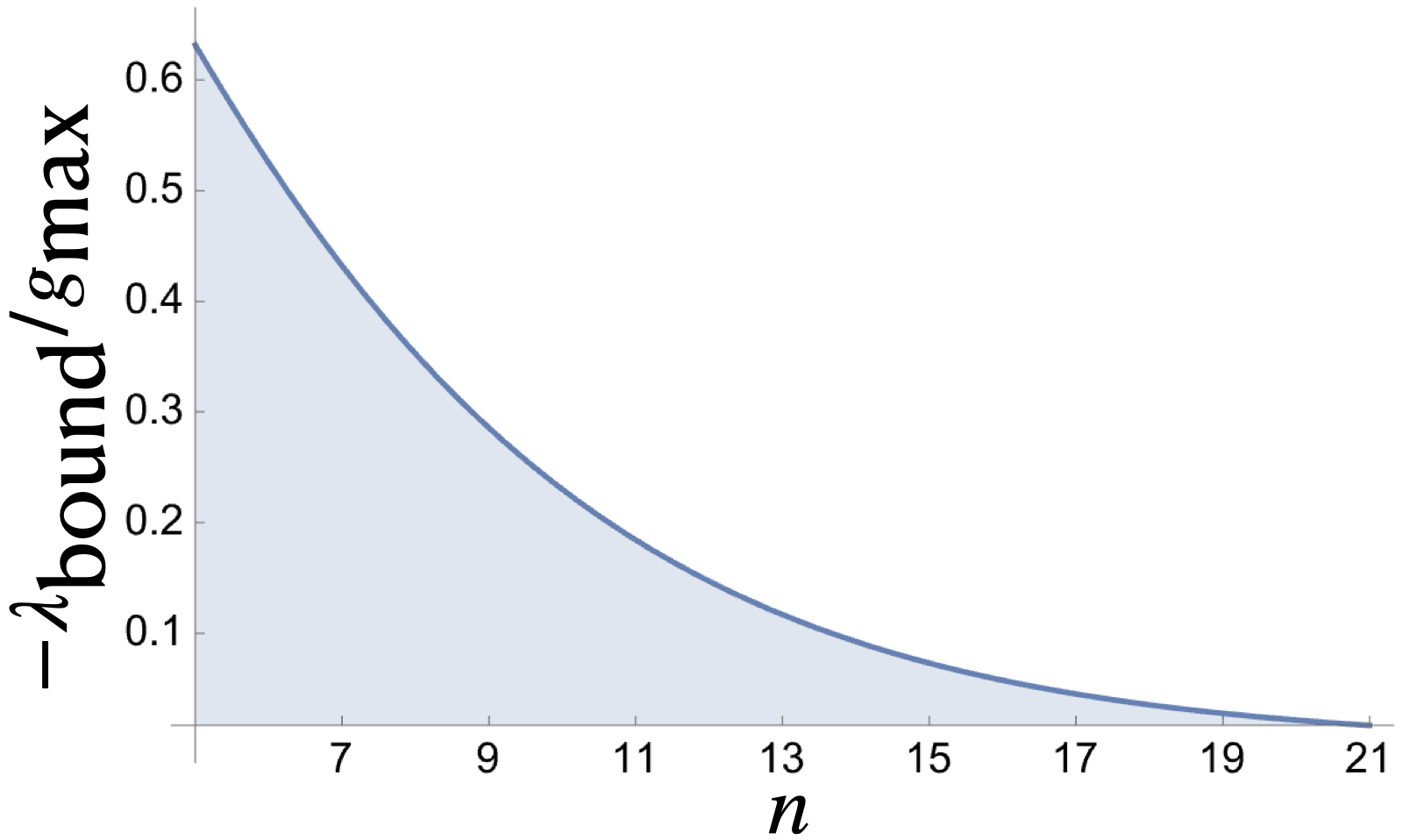}
\caption{Plot of bound (\ref{eq:bound}) on $\lambda_1$ as a function of $n$ for the particular choice $\beta J = 0.8$.  As $n$ increases to $\infty$, the code becomes increasingly stable despite the fact that the temperature $k_B T$ exceeds $J$.}
\label{fig:lambda}
\end{figure}

The thermal behavior of the code implied by the bound (\ref{eq:bound}) is illuminating.  The bound is plotted in Fig. \ref{fig:lambda} as a function of $n$ at $\beta J = 0.8$.  It approaches zero with increasing $n$.  Thus, if there are sufficiently many peripheral qubits on the perimeter, the system becomes stable even if the temperature is relatively large.  

In this sense, there is no phase transition at any finite temperature from a code that can store a logical bit to one that cannot.  The peripheral qubits each possess only a single bond, which might seem reminiscent of a 1-dimensional Ising chain.  Since a 1-dimensional Ising chain is thermally unstable, one might conjecture that the centralized repetition code is also thermally unstable.  However, the connectivity to the central qubit ensures that the system is thermally more similar to a repetition code on a fully connected complete graph, which is stable as shown in the Appendix.  Indeed, changing the logical bit requires passing through high-energy configurations with approximately half the peripheral qubits oriented opposite the central qubit.

It is perhaps instructive to interpret the system's behavior with an analogy to governance: a leader (the central qubit), supported by a majority of subordinates, can steadily guide remaining subordinates.  Alternatively, one can think of a central communication hub that interacts with distant parties, receiving and propagating out information.  The central hub keeps the information relatively robust.  In contrast, the lack of a central hub in a 1-dimensional Ising chain leads to information instability, like a message whispered from party to party in the telephone game.  These perspectives bear a certain relationship to the use of statistical physics to model social dynamics \cite{Castellano2009}.

So far, we have taken the parameter $J$ as a constant independent of $n$.  In many physical contexts, one would expect $J$ to decrease with $n$.  What are the implications for the bound (\ref{eq:bound})?  Suppose that $J(n)$ falls off with $n$ like $\sqrt{\frac{\log n}{n}}$.  Then, we expect $(\cosh \beta J(n))^n$ to be roughly proportional to $n$, implying that the bound (\ref{eq:bound}) approaches a constant with large $n$.  If $J(n)$ falls off slower than this, the bound should approach zero, so that the code can robustly store information.  If $J(n)$ falls off faster than this, the bound should diverge.  It is interesting to note that the borderline case $J \sim \sqrt{\frac{\log n}{n}}$ is similar to $J \sim 1/\sqrt{n}$.  This behavior is suggestive of an arrangement of qubits in 2 dimensions wherein $n$ qubits can be arranged within a distance $\sqrt{n}$ of the hub.  In such an arrangement, if $J$ were to decrease as 1/distance, a falloff $J \sim 1/\sqrt{n}$ might be expected to emerge.

\section{Conclusion}
We have investigated the lifetime of the centralized repetition code coupled to a thermal environment.    The code can robustly store a logical bit against thermal excitation for a time that scales exponentially with the total number of physical qubits $n$. This exponential protection holds for all temperatures $T < \infty$.

Despite the fact that the code has only 1 Ising bond per qubit like a 1-dimensional Ising chain, its star-graph configuration confers thermal stability.  Indeed, it seems likely that the centralized repetition code will store a logical bit longer than any other configuration with 1 Ising bond per qubit.   Still, this noteworthy property cannot be applied naively to the QAC context.  The centralized repetition code does not directly suppress phase errors, which will play a role especially in the first part of a quantum annealing schedule, when there is a transverse Zeeman field present and the energy eigenstates are not of the form $|b_1,\dots,b_n\rangle$.  When such errors are important, further study would be needed to assess the utility of using a large-$n$ centralized repetition code in a quantum annealing device.

\section*{Appendix: Thermal lifetime of the repetition code on a complete graph}

For completeness, this appendix demonstrates that the repetition code on a complete graph is stable, using the same method employed in the body of the paper.  The stabilizer Hamiltonian of the code is
\begin{equation}
H^{\text{cg}} = -J \sum_{1 \le i < j \le n} Z_i Z_j
\end{equation}
with $J>0$.
In terms of the total magnetization $\sum_{i=1}^n (-1)^{b_i}$, the energy eigenvalues can be simplified to
\begin{equation}
E_{b_1,\dots,b_n} = -\frac{J}{2} \left( (\sum_{i=1}^n (-1)^{b_i})^2 - n \right). \label{eq:cgenergy}
\end{equation}

We assume each qubit interacts with the thermal bath as in (\ref{eq:Hint}). The total system evolution is described by the classical master equation $d\mathbf{p}/dt = M^{\text{cg}} \mathbf{p}$, where $M^{\text{cg}} = \sum_{i=1}^n M_i^{\text{cg}}$ incorporates single-qubit transition rates.

The equilibrium Boltzmann distribution $\mathbf{p}_\beta$ for the complete graph has elements
\begin{equation}
p_\beta(b_1,\dots,b_n) = \frac{e^{\frac{\beta J}{2}\left( (\sum_{i=1}^n (-1)^{b_i})^2 - n \right)}}{\sum_{b_1,\dots,b_n} e^{\frac{\beta J}{2}\left( (\sum_{i=1}^n (-1)^{b_i})^2 - n \right)}},
\end{equation}
and satisfies $M^{\text{cg}} \mathbf{p}_\beta = 0$, corresponding to the steady-state eigenvalue $\lambda_0 = 0$. Using the similarity transformation $\tilde{M}_i^{\text{cg}} = P_\beta^{-1/2} M_i^{\text{cg}} P_\beta^{1/2}$, where $P_\beta$ is a diagonal matrix with $\mathbf{p}_\beta$ on the diagonal, we confirm that $\tilde{M}_i^{\text{cg}}$ is Hermitian.  This guarantees that $M_i^{\text{cg}}$ has real eigenvalues. Applying the Gershgorin circle theorem to $M_i^{\text{cg}}$ confirms that all of the eigenvalues $M^{\text{cg}} = \sum_i M_i^{\text{cg}}$ are negative semi-definite.

To bound $\lambda_1$, we construct the variational ansatz $\mathbf{p}^a = \mathcal{Z} \mathbf{p}_\beta$, where $\mathcal{Z}$ is a diagonal matrix with the  logical-Z values $z_{b_1,\dots,b_n} = (-1)^{\text{majority}(b_1,\dots,b_n)}$ on its diagonal. Note that $P_\beta^{-1/2} \mathbf{p}^a$ is orthogonal to $P_\beta^{-1/2} \mathbf{p}_\beta$.  Thus, the following variational argument provides an upper bound on $-\lambda_1$:
\begin{widetext}
\begin{align}
-\lambda_1 & \le - \frac{({\bf p}^a)^T P_\beta^{-1/2} \tilde{M}^{\text{cg}} P_\beta^{-1/2} {\bf p}^a}{({\bf p}^a)^T P_\beta^{-1/2} P_\beta^{-1/2} {\bf p}^a} = - [1,\dots,1] {\cal Z} M^{\text{cg}} {\cal Z} {\bf p}_\beta = 4 [1 \, \dots \, 1] \left(\frac{I-{\cal Z}}{2}\right) \left(\sum_{i=1}^n M_i^{\text{cg}}\right) \left(\frac{I+{\cal Z}}{2}\right) {\bf p}_\beta \nonumber \\
& = 4 \frac{\sum_{i=1}^n \sum_{b_1,\dots,b_{i-1},b_{i+1},\dots,b_n  \text{ half zero}} g_i(E_{b_1,\dots,b_{i}=0,\dots,b_n} -E_{b_1,\dots,b_{i}=1,\dots,b_n}) e^{-\beta E_{b_1,\dots,b_{i}=0,\dots,b_n} }}{\sum_{b_1,\dots,b_n} e^{-\beta E_{b_1,\dots,b_n}}} \nonumber \\
& = 4 \frac{\sum_{i=1}^n g_i(0) \binom{n-1}{\frac{n-1}{2}} e^{\frac{\beta J (n-1)}{2}}}{\sum_{b_1,\dots,b_n} e^{-\beta E_{b_1,\dots,b_n}}} \le 2 g_{\text{max}} n \binom{n-1}{\frac{n-1}{2}} e^{-\frac{\beta J}{2} (n-1)^2} = -\lambda_{\text{bound}}^{\text{cg}}
\label{eq:bound_complete}
\end{align}
\end{widetext}
where $g_{\text{max}} = \max_{i} g_i(0)$.   In this argument, the matrix $\left(\frac{I-\mathcal{Z}}{2}\right) M_i^{\text{cg}} \left(\frac{I+\mathcal{Z}}{2}\right)$ is non-zero only for transitions from configurations with logical bit value 0 to configurations with logical bit value 1.  Thus, for an odd number of qubits $n$, if qubit $i$ is undergoing a bit flip, the remaining $n-1$ qubits must be split evenly with $(n-1)/2$ zeros and $(n-1)/2$ ones.  Flipping the $i$-th qubit in such a configuration does not change its energy; equation (\ref{eq:cgenergy}) evaluates to $J/2(n-1)$ for both the initial and final states.  As a result, the transition rate is simply $g_i(0)$.  Approximating the partition function in the denominator of (\ref{eq:bound_complete}) by its ground-state contributions $2 e^{\frac{\beta J (n^2-n)}{2}}$ leads to the final bound.

Note that $\lambda_{\text{bound}}^{\text{cg}}$ decays quadratically-exponentially in $n$ so that the logical bit lifetime $\tau \sim 1/|\lambda_1|$ is extremely long.  Indeed, even if one assumes that the interactions decay with $n$ as $J =  \jmath/n$, the bound (\ref{eq:bound_complete}) still implies an exponentially growing lifetime provided that $\beta \jmath \gtrsim 2 \ln 2$.

%Just as in the star-graph case, the error in the variational guess $1 - |a_1|^2$ can be bounded by establishing a finite spectral gap $\lambda_2 \le -G^{\text{cg}}$ for bulk relaxation modes, confirming that $\mathbf{p}^a$ converges exponentially fast to the true second eigenvector as $n \to \infty$.

\bibliography{references.bib}

\end{document}